\documentclass[preprint,tightenlines,showpacs,nofootinbib,aps,prc]{revtex4} 

\usepackage{graphicx}% Include figure files
\usepackage{bm}% bold math
\usepackage{url}
\def\be{\begin{equation}}
\def\ee{\end{equation}}
\def\twop{$2^+$~}
\def\lb{\langle}
\def\rb{\rangle}
\def\pb{$^{208}$Pb}
\begin{document}

\title{Systematics of the first $2^+$ excitation in spherical nuclei with 
Skyrme-QRPA}

\author{J.~Terasaki}
 \affiliation{Department of Physics and Astronomy, University of North Carolina,
Chapel Hill, NC 27599-3255}
\author{J.~Engel}
 \affiliation{Department of Physics and Astronomy, University of North Carolina,
Chapel Hill, NC 27599-3255}
\author{G.F.~Bertsch}
\affiliation{Department of Physics and Institute for Nuclear Theory,
University of Washington, Seattle, WA 98195}
\def\skm{SkM$^*$~}

\begin{abstract}
We use the Quasiparticle Random Phase Approximation (QRPA) and the Skyrme
interactions SLy4 and \skm to systematically
calculate energies and transition strengths for the lowest $2^+$
state in spherical even-even nuclei.
The \skm functional, applied to 178 spherical nuclei between $Z=10$ and 90,
produces excitation energies that are on average 11\% higher than
experimental values, with residuals that fluctuate about the average by 
$-35\%+55\%$.  The predictions of \skm and
SLy4 have significant differences, in part because of differences in the calculated ground state
deformations;  \skm performs better 
in both
the average and dispersion of energies.  
Comparing the QRPA results with those of
generator-coordinate-method (GCM) calculations, we find that the QRPA reproduces
trends near closed shells better than the GCM, and overpredicts the energies
less severely in general.  We attribute part of the difference to a deficiency
in the way the GCM is implemented.
\end{abstract}

\maketitle
\section{Introduction}

Computer resources now allow nuclear Density Functional Theory (DFT), also
called self-consistent mean field theory, to be applied systematically over the
entire nuclear chart.  Although it is usually applied to ground states, static
DFT can be extended to treat excitations.  A number of methods to do so have
been developed, and it is not yet clear how accurate any of them is and how they
compare to one another.  In this paper we apply one of the methods, the
Quasiparticle Random Phase Approximation (QRPA), to calculate the properties of
the first excited \twop states in even-even nuclei. We assess strengths and
weaknesses of the method for spherical nuclei and compare results with those of
two other systematic studies that used different methods.  We have previously
carried out QRPA calculations for several isotope chains leading to the neutron
drip line \cite{Ter06}; the same theory and codes are used here to cover
essentially all known lowest-lying \twop states in spherical nuclei.
 
Among the extensions of static DFT, the QRPA is one of the best justified
theoretically.  It is the adiabatic limit of time-dependent DFT and as such may
be derived from a variational principle.  Time-dependent DFT preserves dynamical
conservation laws, and in the adiabatic approximation is unambiguously defined
by the density functional used for the static properties.  Its major shortcoming
as an approximation to many-body dynamics is that it works reliably only when
ground state is nearly a single Slater determinant or a BCS-like condensate.

Recently, two other systematic calculations of \twop excitations that use very
different approximation schemes \cite{Sab07,Gou07} have appeared.  Both of these
studies constrain DFT by applying external quadrupole fields.  This allows one
to build a multi-configurational wave function or construct a collective
potential-energy surface in deformation space.  Reference \cite{Sab07}, using
the SLy4 energy-density functional \cite{Cha98}, treats
many constrained configurations together to make a discrete-basis Hill-Wheeler
approximation.  Reference \cite{Gou07}, using the Gogny interaction \cite{Dec80}
constructs a collective Hamiltonian in
the 5-dimensional quadrupole-deformation space via the collective potential
energy surface and a local treatment of the kinetic energy operator.  We will
refer to both as Generator-Coordinate Method (GCM) calculations.
 
In addition to the approximation scheme, the
quality of the energy-density functional also affects accuracy; most
functionals are constructed to optimize their static properties.  In this study
we cannot hope to make better functional, but we can at least compare the
quality of those that already exist.  Here we will test two Skyrme functionals,
SLy4 \cite{Cha98}, and SkM$^*$ \cite{Bar82}, both in conjunction with ``volume"
pairing interactions. The results for SLy4 can be directly compared to those of
Ref.\ \cite{Sab07}, which uses the same functional\footnote{In our previous
work we also considered the SkP functional \cite{Do84}, but found that under
certain conditions it lacks a true variational minimum \cite{le06}.}.  We should
note that a relativistic functional has already been applied in the QRPA to a subset of the nuclei we
examine here \cite{an06}.

\section{Implementation of QRPA}

Our QRPA and underlying Hartree-Fock-Bogoliubov (HFB) calculations assume spherical symmetry and
employ a radial mesh with box boundary conditions.  The HFB orbitals are
calculated with a code based on the one
described in Ref.  \cite{Do84}.  For the global survey here, we give the
box a radius of 16 fm and a radial mesh size of 0.05 fm.  Although we can take
the ordinary DFT functional from mass fits in the literature, the choice of
a pairing functional is less clear cut.
We have chosen to treat  pairing through
an ordinary contact interaction, $-V_0\,\delta(\mathbf{r}_1-\mathbf{r}_2)$.
After truncating the space of single-particle orbits (see below), we  make an
approximate fit of the pairing strength to experimental data.  The quantities
compared are the average HFB pairing gap \cite{do96} and the 3-point difference
$\Delta^{(3)}$ of the experimental binding energies.  The estimated strengths
with the SLy4 functional are $V_0=200$ MeV fm$^3$ for neutrons and
$V_0=240$ MeV fm$^3$ for protons
\footnote{The finding that the extracted proton pairing strength is
larger than the neutron value appears in other studies
as well, e.g.\ Ref.\ \cite{go02}. The strengths for the SkM$^*$ functional are
$V_0= 170$ MeV-fm${^3}$ for neutrons and $V_0= 200$ MeV-fm${^3}$ for
protons.}.

We use the matrix representation of the QRPA in a two-quasiparticle basis, as
described in detail in Ref.\ \cite{Ter05}.  We represent the HFB eigenfunctions
in the canonical basis, and truncate them through limits on the occupation number
($v^2$) or, if the pairing gap is zero, on the single-quasiparticle energy \cite{Ter05}.
We then construct the usual $A$
and $B$ matrices \cite{Rin80} in the truncated two-quasiparticle basis.  

The method was applied to the Ca, Ni, and Sn isotopes for $0^+$, $1^-$, and
$2^+$ states in Ref.\ \cite{Ter06}.  In this paper we apply the QRPA to many
more nuclei than in Ref.\ \cite{Ter06}, but only consider the low-lying $2^+$
states.  Unlike the  $0^+$ and $1^-$ excitations, the $2^+$ states are not
affected by spurious modes in spherical nuclei; we can therefore afford to use a
smaller space.  In our prior work, for example, we used a box radius of at least
20 fm. Here we reduce the radius because the lowest $2^+$ states are well
localized, even near the drip line.  Because we do not have to be as careful
about translational symmetry, we reduce the quasiparticle-energy cutoff in the
HFB calculation from 200 MeV to 50 MeV.  Finally, we increase the
canonical-basis occupation-number cutoff --- the smallest occupation number that
canonical states included in the QRPA can have --- from $10^{-8}$ to $10^{-7}$,
and reduce the single-particle-energy cutoffs (used when pairing is absent) from
100 MeV to 30 MeV.  The only way in which we must extend the basis
is by increasing the cutoff in the single-particle angular momentum
from 21/2 (in our previous work) to 25/2.  The reason is that we are dealing
with heavier nuclei here.  With all of these limits in place, our largest QRPA
matrix, which we encountered in the very heavy nucleus $^{220}$Th, has dimension
6086.  The space defined by all the cutoffs is large enough so that energies and
$B(E2)\!\!\uparrow$ strengths do not change when we make it larger still.  With the
space size fixed, we then determine the strength $V_0$ of the pairing
interaction as described above.

\section{Selection of nuclei and their QRPA eigenmodes}

Our goal is to evaluate the performance of the QRPA in all the spherical nuclei
with $10 \leq Z \leq 90$ for which experimental data have been tabulated in
Ref.\ \cite{Ram01}.  That reference reports 507 nuclei with measured 2$^+$
excited states.  We select spherical nuclei as follows:
We use the HFB
code ev8 \cite{ev8} to do constrained calculations on a grid of deformations,
and drop the cases in which the lowest-energy configuration has nonzero
deformation.  For the remainder, we perform unconstrained calculations, starting
from deformation values Q= 0 and $\pm 200 $ fm$^2$.   If any of these
configurations evolves to a lower energy than the spherical one, we drop the
nucleus.  

A final requirement for selection is that the QRPA
eigenvector correspond to a physical excitation.  Thus, we
exclude two nuclei in the SLy4 data because the eigenvalues are
imaginary\footnote{In principle, the stability of the spherical ground state
demands that the eigenvalues all be real.  However, the difference between the
ways we truncate the HFB and QRPA calculations gives rise to
occasional violations.}.
Other
unphysical excitations are eigenvectors whose predominant components correspond
to pair addition or removal, rather than excitation of the ground state.  These
components are present because the HFB quasiparticle vacuum does
not have a well-defined number of particles.  We
exclude eigenvectors that have nearly pure two-particle-transfer character,
using as a measure an approximate difference in particle number between the ground
and excited states:
\begin{equation} \Delta N = 2 \sum_{ij} (X_{ij}^2-Y_{ij}^2)(1-v_i^2-v_j^2)\,.
\label{eq:DN} 
\end{equation}
Here $v^2$ is a canonical-basis condensate
occupation probability, $i,j$ label the quasiparticles in the two-quasiparticle
basis, and $X,Y$ are the usual QRPA amplitudes.  If $|\Delta N|$ is large, we
select another excitation or drop the nucleus from the table altogether.  The
final tables contain 155 spherical nuclei for the SLy4 functional and 178 for
the SkM* functional.  We note that the two tables have 129 out of a possible 155
nuclei in common.  The imperfect overlap indicates that sphericity is not a very robust property of the
DFT, unlike strong deformation.   The data sets are posted with the calculated
excitation properties in Ref.\ \cite{epap}.    

\begin{figure}[t]
\includegraphics[width=8cm]{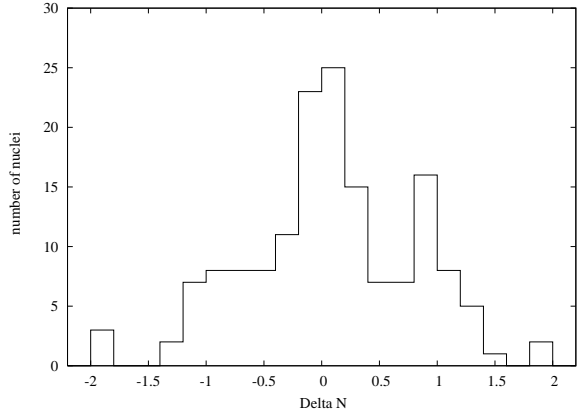}
\caption{\label{fig:deltaN} Particle-hole character of the 
lowest \twop solutions.  The histogram displays the quantity 
$\Delta N$ defined in Eq. (1) for 155 nuclei in the SLy4 data set 
(one of which we drop --- see text).  The values $-2,0,+2$ correspond to excitations
of hole-hole, particle-hole, and particle-particle character, respectively.}
\end{figure}

Figure \ref{fig:deltaN} shows a histogram of values of $\Delta N$
for the SLy4
data set.  There is a well-defined peak at $\Delta N =0$ with
a spread of $\pm 0.4$.  Beyond that, a plateau covers the
range $\pm 1.3$.  All these nuclei are included in the data set.  
Finally, there are 5 transitions that are clearly
at the two-particle transfer limit $\Delta N \approx \pm 2$.  These
nuclei are $^{40,48}$Ca, $^{68}$Ni, $^{80}$Zr, and $^{132}$Sn.  For
all but one ($^{40}$Ca), an acceptable transition is present at
an energy slightly higher than that of the lowest-energy solution
and the nucleus is included in our survey.

\begin{figure}[t]
\includegraphics[width=11cm]{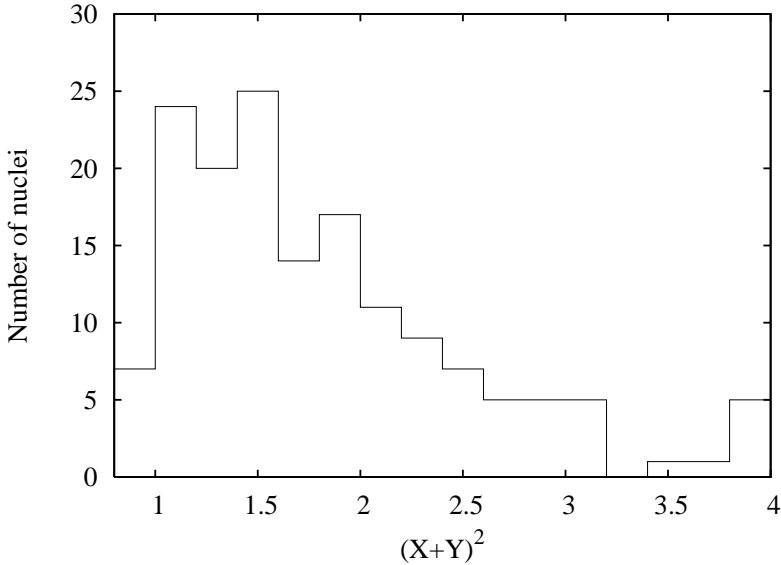}
\caption{\label{fig:x+y} Distribution of the summed $(X+Y)^2$
amplitudes, Eq.~(\ref{eq:xpy}), for the first \twop transitions in spherical nuclei,for 
the SLy4 functional. Values greater than 4 are consolidated in the
last bin.   
}
%see gene:terasaki/ph-pp/ histogram_xpy.py, xpx_hist,gnu
\end{figure}

We will also want to examine how well the QRPA handles ``soft''
nuclei, i.e.\ those which
make large excursions from the sphericity.
The size of the $Y$ amplitudes is a good measure of softness.  The
enhancement of $T$-even transition rates due to the added ground
state QRPA correlations is given roughly by the softness parameter
\begin{equation}
\label{eq:xpy}
C=\sum_{ij} (X_{ij}+Y_{ij})^2. 
\end{equation}
This expression is designed to reduce to unity in the absence of 
the $Y$ amplitudes and have the basic $(X+Y)^2$ structure 
characteristic of QRPA transition rates.
A histogram for $C$ appears in Figure \ref{fig:x+y}.  
The softness factors range from 1 (no
softness) to more than 4.  Even the values in the middle of the
histogram would make the Quasiparticle Tamm-Dancoff Approximation
invalid.  The QRPA itself can have problems with very soft nuclei; if the $Y$ amplitudes are comparable in 
size with the $X$ amplitudes, then a phase transition to a ground-state of
totally different character is nearby.

\section{Energies}

\begin{figure}[t]
\includegraphics[width=11cm]{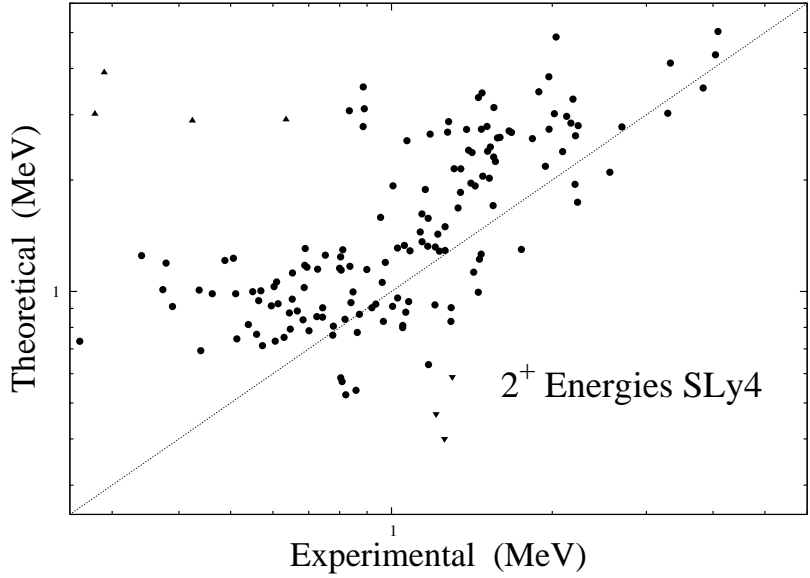}
\caption{\label{fig:e_sly4} Calculated energies for lowest
$2^+$ states in spherical nuclei, plotted versus experimental energies.  The
theoretical energies are from QRPA with the SLy4 energy functional.
The experimental data are from Ref.\ \cite{Ram01}. For triangles, see text.
}
% gene:terasaki/e_compare*
\end{figure}

Figure \ref{fig:e_sly4} shows our predicted energies for the lowest \twop states
with
the SLy4 functional; the predictions are plotted versus
measured energies.
The data span a range of 
more than an order of magnitude and the theory captures much of 
that variation.  There are some cases where
the errors are very large, however.  Far off the line on the upper left
hand side are nuclei with $N=40, Z=34-40$, shown by triangles.
They are predicted to be spherical by the SLy4 functional
but experimentally they appear to be deformed.  On the lower side
of the diagonal, the outlying cases are the isotopes of Sn ($Z=50$) with
neutron numbers $N=60,62,64$ (inverted triangles).  In this case, the SLy4 functional 
predicts the nuclei to be very soft with respect to quadrupole 
deformation.   In fact, these 3 nuclei have the highest values of 
the softness parameter in the data set.

\begin{figure}[t]
\includegraphics[width=.7\textwidth]{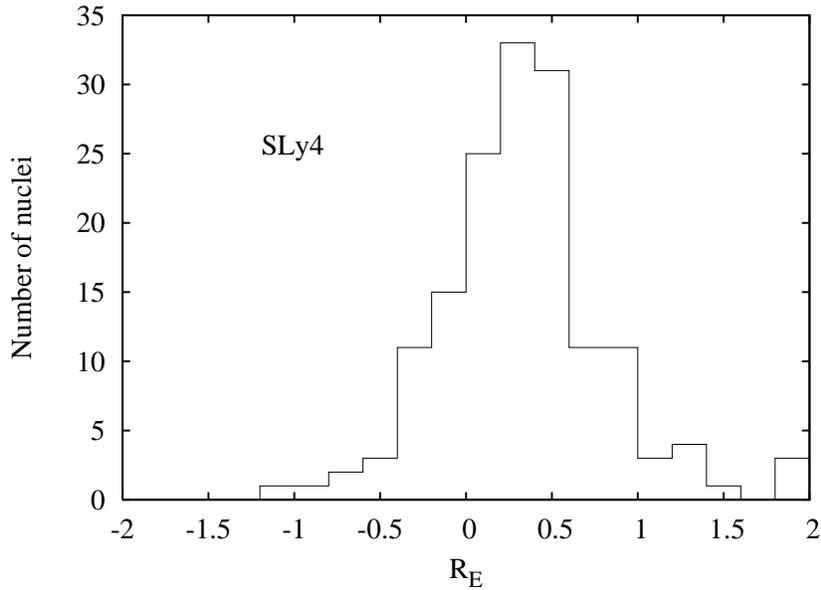}
\caption{\label{fig:ehist} Histogram of the quantity $R_E$ for the 155
nuclei in the SLy4 data set. The highest bin includes 3 nuclei having
$R_E > 2 $.  
}
\end{figure}

We can 
quantify the performance of calculations through the measure
\begin{equation}
R_E\equiv
\textrm{ln}(E_\mathrm{calc.}/E_\mathrm{exp.})\,,
\end{equation}
as is done in Ref.\ \cite{Sab07}.   Figure \ref{fig:ehist} shows a histogram
of $R_E$ for the 155 nuclei in the SLy4 data set.
% gene:terasaki/ e_hist.*
%
We see that theory tends to overpredict the excitation
energy, but the error is systematic and the overall distribution
is strongly peaked.  To summarize performance in a single number, we can use
the average $\bar R_E$ for the
data set.  For spherical SLy4 nuclei, $\bar R_E = 0.33$, corresponding
to a calculated energy about 40\% higher than the experimental value.  Another
indicator of theory's performance is the width of the peak in the 
histogram.  We define the dispersion of $R_E$ about the average as
\be
\sigma_E = \sqrt{ \lb R_E^2\rb -\bar R^2_E }\,.
\ee
A small value of $\sigma_E$ implies that the theory tracks fluctuations
in the data (though the average is not necessarily correct).  For SLy4, the
dispersion is 
$\sigma_E=0.51$,
corresponding to error bars of $+66/-40$ \% in the energies.  It should
be noted that the peak is far from Gaussian, and the bounds $\pm\sigma_E$
include more of the data set than would be the case for a Gaussian
distribution.  If we were to use the fraction of nuclei inside the bounds to 
define $\sigma_E$, its value would be 20\% lower.   
\begin{table}[b]
\caption{\label{tab:ecompare}
Averages  $\overline{R}_E$ 
and standard deviations $\sigma_E$ for measured
\twop excitations with the SLy4 and
\skm functionals and various cuts on the data set. The set labeled ``common"
consists of the nuclei that are spherical for both the SLy4 and the \skm
functionals.
}
\begin{ruledtabular}
\begin{tabular}{lrccc}
functional    & data set & Number of nuclei&  $\overline{R}_E$ &  $\sigma_E$\\
\hline
SLy4   & all spherical & 155 &0.33 & 0.51\\
        & low $|\Delta N|$ &79 &0.29 & 0.47\\  % DNgood.dat 
& high $|\Delta N|$ & 78& 0.38& 0.54 \\    % DNbad.dat
& low softness  & 106 &0.47 & 0.48\\    % xpy_low.dat
& high softness  & 49 & 0.04 & 0.44\\    % xpy_high.dat
& common  &  129 &0.26 &0.40 \\
\hline
SkM$^*$ & all spherical & 178&  0.11 & 0.44 \\
        & low softness &  115 & 0.27 & 0.35 \\
        & high softness &  63 & $-$0.17 & 0.45 \\
        & common &  129 & 0.14 & 0.38 \\
\end{tabular}
\end{ruledtabular}
% gene:terasaki/ divide_xpy.py statistics3.py e_statistics.py e_all.py
\end{table}

It is interesting to see whether the accuracy of the energy prediction depends
on other characteristics of the state.
To examine this question, we use the value $|\Delta N|$ to split the data set
into two parts.   Dividing the data set at 
$|\Delta N|=0.5$ gives two roughly equal-size subsets.  Table I shows the performance
measures for the two subsets, and
for reference the combined results discussed in the last paragraph.
%gene:terasaki/statistics3.py with DNgood.dat, DNbad.dat
Performance in the high-$|\Delta N|$ set is slightly poorer for both the average and the
dispersion, but in our opinion, the differences are not large enough to warrant 
the use of
$|\Delta N|$ as a selection criterion beyond the most extreme cases.  
%We now say earlier that the Y's are almost never big enough to invalidate the
%QRPAThe accuracy of the QRPA
We also checked the dependence of accuracy on softness, dividing the data into
subsets with $C$ greater and less than 2.  The results appear in lines 4 and 5 of the table.  The separation 
strongly affects the averages, so that the high-softness set is actually more accurately
described, but the dispersions are not significantly
different.  It is not totally surprising that states in soft nuclei are better
described, as long as the nuclei are not too soft, because such states tend also
to be more collective and RPA methods
capture collectivity well.

\begin{figure}[b]
\includegraphics[width=11cm]{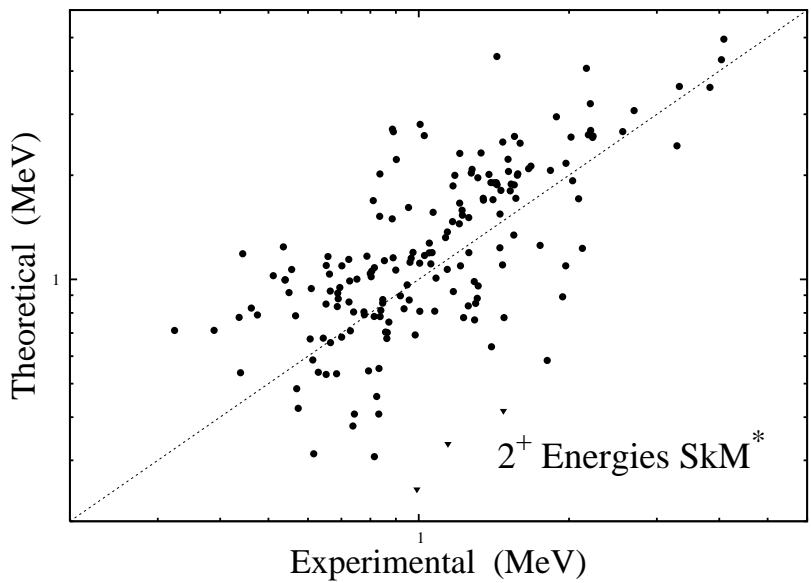}
\caption{\label{fig:e_skm} Same as Figure \ref{fig:e_sly4}, using the \skm functional for
the calculated energies.
}
%gene:terasaki/e_compare2.*
\end{figure}

We next  examine the performance of the \skm functional.  Here the protocol for
selecting spherical nuclei gives a data set with 178 members.  The set
does not include the nuclei near $^{80}$Zr whose energies were poorly predicted
by the SLy4 functional.  Figure \ref{fig:e_skm} shows the scatter plot of theoretical versus 
experimental excitation energies.  There is 
more clustering along the diagonal than for SLy4, but outliers still exist
on the lower side of the plot.  These are all open-shell nuclei, and it is not as easy
to characterize them as it was for SLy4.
The worst cases are the nuclei $^{44}$Ar, $^{64}$Zn, and $^{216}$Th, shown
as the inverted triangles in the figure. What they have in common is
a very high QRPA softness, in the range 
$C=2-8 $.

The $R$ metrics for the \skm data set are given on line 7 of the Table. 
The average is much better (lower both in the high- and low-softness set) than with SLy4, and there is also some
improvement in the dispersion. 
% gene:terasaki/ statistics.py skm_e.dat 
Much of the improvement
comes undoubtedly from the better selection of spherical nuclei, but part may
be due to differences in the effective QRPA interaction that comes from
the functional.  One way to test that is to compare the performance on
the 129 nuclei that the two data sets have in common.  The results are labeled
``common" in Table I.  The averages
still differ significantly --- $\bar R_E({\rm SLy4}) -\bar R_E({\rm SkM}^*)=
0.12$ --- indicating that the functional plays a role beyond merely shaping
the ground state.  It remains to characterize the
aspects of the functional that are responsible, but this is beyond the 
scope of our project here.

\begin{figure}[t]
\includegraphics[width=8cm]{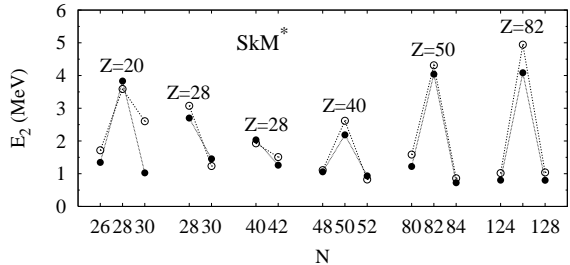}
\caption{\label{fig:dblm} \twop excitation energies in the vicinity of 
doubly magic nuclei.  Displayed
on either side of the doubly magic nucleus $(N,Z)$ are the adjacent
even-even isotopes, $(N-2,Z)$ and $(N+2,Z)$.  Theoretical
energies (open circles) were calculated with the \skm
functional.
Filled circles are experimental energies. 
}
%gene:terasaki/chains/dblm*
\end{figure}

\begin{figure}[t]
\includegraphics[width=8cm]{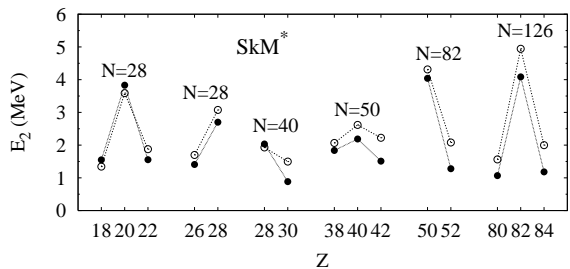}
\caption{\label{fig:dblmZ} Same as Figure \ref{fig:dblm} for 
isotones $(N,Z-2)$, $(N,Z+2)$ adjacent to doubly magic nuclei $(N,Z)$.
}
%gene:terasaki/chains/sn_qrpa*
\end{figure}

The small-amplitude assumption of QRPA is best justified in magic
and semimagic nuclei.   Figures   
\ref{fig:dblm} and \ref{fig:dblmZ} show the very accurate QRPA results with \skm in the
vicinity of doubly magic nuclei in several isotope and isotone chains.  Going from left to right in the figures,
the magic numbers are $(N,Z) = (28,20), (28,28),
(40,28), (50,40), (82,50)$, and $(126,82)$. This last doubly magic nucleus
is the one with the highest \twop energy in all the chains.  We obtain
accurate results around doubly magic nuclei with SLy4 as well.  


\section{Transition strengths}

The $0^+ \rightarrow 2^+$ transition strength for 88 of the 155 nuclei in the
SLy4 spherical data set
have been measured, and are compiled in Ref.\ \cite{Ram01}.  
\begin{figure}[b]
\includegraphics[width=.9\textwidth]{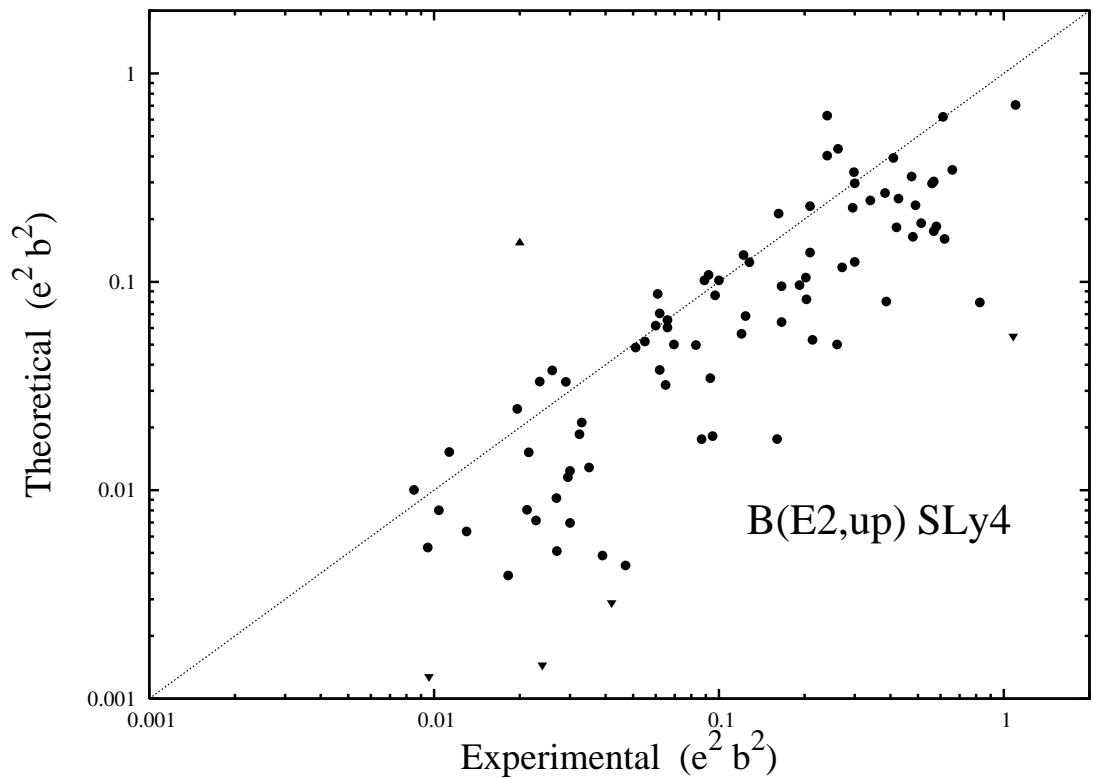}
\caption{\label{fig:be2} The same as Figure~\ref{fig:e_sly4} but for
$B(E2)\!\!\uparrow$. 
}
%gene:terasaki/q_compare.py    *high.dat, *low.dat, *trunc.dat
\end{figure}
Figure \ref{fig:be2} shows a scatter plot comparing the 
SLy4 predictions of $B(E2)\!\!\uparrow$ with the experimental numbers.
Four nuclei for which the theory seriously underpredicts the data 
are marked by triangles on the lower side of the scatter plot.  The 
nuclei are $^{38}$Ca, $^{34}$Ar, $^{42}$Ca, and $^{78}$Sr, going from
left to right on the plot.  The nucleus $^{78}$Sr is in the
deformed $(N,Z)\sim (40,40)$ region; its large experimental
transition strength confirms its
deformed character.  The failure of the theory for the Ca isotopes 
near $N=20$ can be explained by the absence in our calculation of deformed
intruder orbits. 
Only one nucleus on the list,
$^{34}$Ar, does not have any obvious properties that would cause
the QRPA to fail.

On the upper side of the plot, we have marked the most prominent 
outlier, 
the nucleus $^{210}$Po.  Its measured strength is nearly an order of
magnitude smaller than the calculated values for both the SLy4 and the \skm
functionals.  This discrepancy is one of the most puzzling in the
survey, because deformation effects would only increase the transition strength.
The predominant amplitude in the transition, with 93\% of the 
normalization sum $\sum X^2-Y^2 = 1$, has both quasiparticles in
the lowest $h_{9/2}$ states, corresponding to the shell-model 
transition $(h_{9/2}\,h_{9/2})^{J=0} \rightarrow (h_{9/2}\,h_{9/2})^{J=2}$.
By itself, the pure shell-model transition has a strength of
454 e$^2$fm$^4$.  The additional small $X$ amplitudes\footnote{
The $Y$ amplitudes play a
minor role; the softness parameter C in Eq. (\ref{eq:xpy}) is 
close to unity ($C= 1.09$).} in the QRPA 
increase the strength by factor of 3.4. An
enhancement of the single-particle strength, often represented by 
effective shell-model charges, is nearly universal in quadrupole
transitions. 
The nucleus $^{210}$Po is an exception,
with a measured value 200 e$^2$fm$^4$, even smaller than the
shell-model result with bare charges.  
	
To summarize the performance of the QRPA for transition strengths,
we follow Ref.\ \cite{Sab07} and
define the residual of the transition matrix element logarithm $R_Q$:
\begin{equation}
R_Q\equiv
\textrm{ln}\left[\sqrt{B(E2)\!\uparrow_\mathrm{calc.}/
B(E2)\!\uparrow_\mathrm{exp.}}\right]\,.
\end{equation}
\begin{figure}
\includegraphics[width=.9\textwidth]{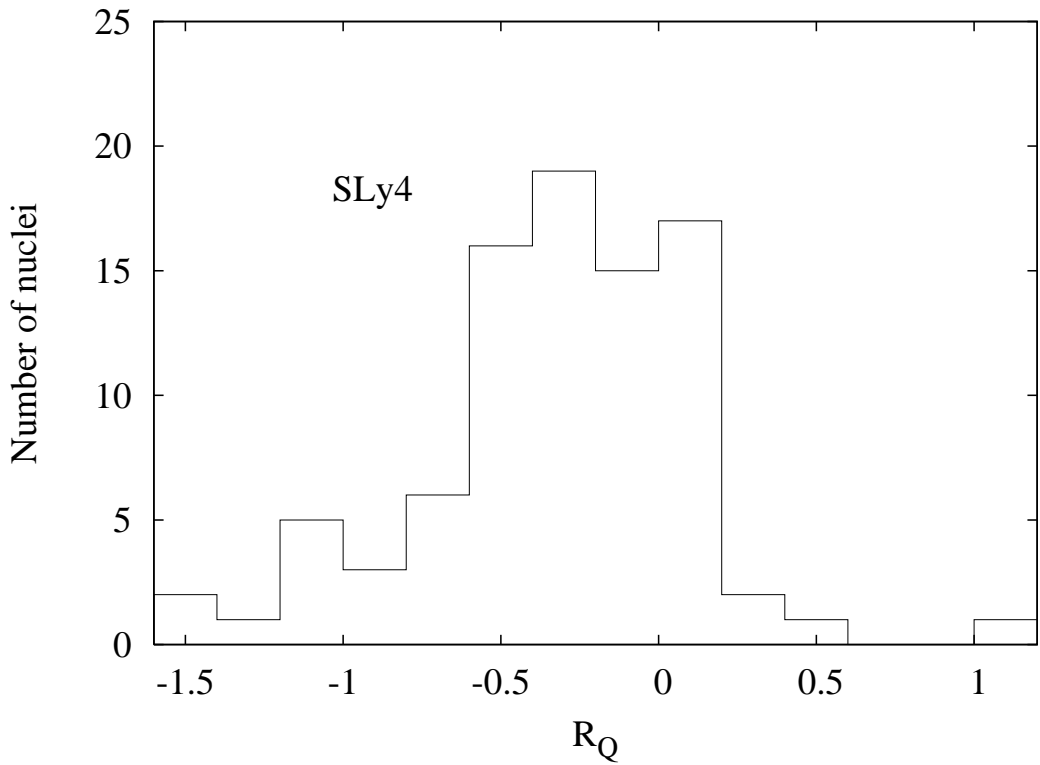}
\caption{\label{fig:qhist} Histogram of $R_Q$ for the 155
nuclei in the SLy4 data set. The highest bin represents the nucleus
$^{210}$Po.
}
\end{figure}
% gene:terasaki/ q_hist.*
%
Figure \ref{fig:qhist} shows a histogram of this quantity for the SLy4
functional.  $^{210}$Po on the right clearly stands out as 
an anomalous case.  The long tail on the left represents mainly 
deformed nuclei that are incorrectly predicted to be spherical.
The average and
standard deviations of $R_Q$ for both functionals are given in 
Table \ref{tab:qstat}.  There is very little 
difference between the performance of the two functionals when it comes to
transitions.

\begin{table}
\caption{\label{tab:qstat}
Averages  $\overline{R}_Q$ and
standard deviations $\sigma_Q$ of the
distributions in $R_Q$ for the SLy4 and SkM$^*$ functionals.
}
\begin{ruledtabular}
\begin{tabular}{lrcc}
 functional   &  $\overline{R}_Q$ & $\sigma_Q$\\
\hline
SLy4   & -0.32 & 0.42\\
SkM$^*$  & -0.29 & 0.53\\
\end{tabular}
\end{ruledtabular}
%gene:terasaki/q_all.py, e_statistics.py
\end{table}

\section{Comparison with GCM calculations}

We refer to the GCM-based techniques of Refs.\ \cite{Sab07} and \cite{Gou07} as
GCM-Hill-Wheeler (GCM-HW) and GCM-5-Dimensional-Collective-Hamiltonian
(GCM-5DCH) respectively.  The GCM-HW study is in some ways the most like ours.
It used the SLy4 functional, making the results directly comparable to our SLy4
results.  And unlike the GCM-5DCH, it was applied to doubly magic nuclei.  Its
set of spherical nuclei is somewhat different from ours because of differing
pairing interactions, so we use overlap of the two sets for comparison.  

The results are in Table III.   The energy measures show some interesting
differences.  From the $R_E$ values we see that the GCM-HW gives energies about
40\% higher than those of the QRPA.  A possible explanation for the
overprediction of energies by the GCM-HW in spherical nuclei appears in the
Appendix A.  The other interesting difference, which we cannot explain, is that
the GCM-HW fluctuations track the experimental ones better than the QRPA.

The $B(E2)$$\uparrow$ metrics appear in the last two columns
of the table.  While the QRPA seriously underpredicts the transition
strengths, the
GCM-HW errs in the opposite direction.  It is not surprising that
the GCM wave function produces larger quadrupole matrix elements
because it incorporates large deformation (unlike the QRPA), and because the $J=0$ and $J=2$
states have a similar intrinsic structure.  
\begin{table}
\caption{\label{tab:e-compare}
Averages  $\overline{R}_E$ 
and standard deviations $\sigma_E$ for 153 
\twop excitations produced with the SLy4 functional by our QRPA
calculations and the GCM-HW calculations \cite{Sab07}.
}
\begin{ruledtabular}
\begin{tabular}{lcccc}
theory    &   $\overline{R}_E$ &  $\sigma_E$&$\overline{R}_Q$ &  $\sigma_Q$\\
\hline
QRPA &  0.33 & 0.51 & -0.32 & 0.42\\
GCM-HW    &   0.67 & 0.33 & 0.16 & 0.41 \\  
\end{tabular}
\end{ruledtabular}
% gene:terasaki/statistics.py, q_statistics.py
% 2+_all.dat from gene:/scratch/2+/tables
% gene:terasaki/read_sabbey.py, read_sabbey_be2.py
%   -> map_e.dat, hw_e.dat, map_be2.dat,hw_be2.dat
% map_e:  153 0.61 0.48
\end{table}

Unfortunately, because Ref.\ \cite{Gou07} used a different functional, it is more
difficult to compare with our study; it is not
possible to distinguish effects of the functional from those of
the methodologies.  We compare results nevertheless. 
As before, we use spherical nuclei analyzed in both studies to 
compute the $R$-statistics; 
doubly magic nuclei are thereby excluded.  We show in the first line of 
Table \ref{tab:skm-goutte}  the $R$-statistics
of the QRPA with our better-performing functional, SkM$^*$;
\begin{table}
\caption{\label{tab:skm-goutte}
Same as Table \ref{tab:e-compare}, with the SkM$^*$ functional for 
the QRPA and the Gogny functional for the GCM-5DCH.
}
\begin{ruledtabular}
\begin{tabular}{lcccc}
theory    &   $\overline{R}_E$ &  $\sigma_E$&$\overline{R}_Q$ &  $\sigma_Q$\\
\hline
QRPA (SkM$^*$)&  0.10 & 0.45 & -0.29 & 0.51\\
GCM-5DCH (Gogny)    &   0.19 & 0.43 & 0.22 & 0.27 \\  
\end{tabular}
\end{ruledtabular}
% gene:terasaki/statistics.py, q_statistics.py
% read_goutte.py to produce goutte_e.dat, goutte_be2.dat
\end{table}
\begin{figure}[b]
\begin{flushleft}
\includegraphics[height=4.3cm]{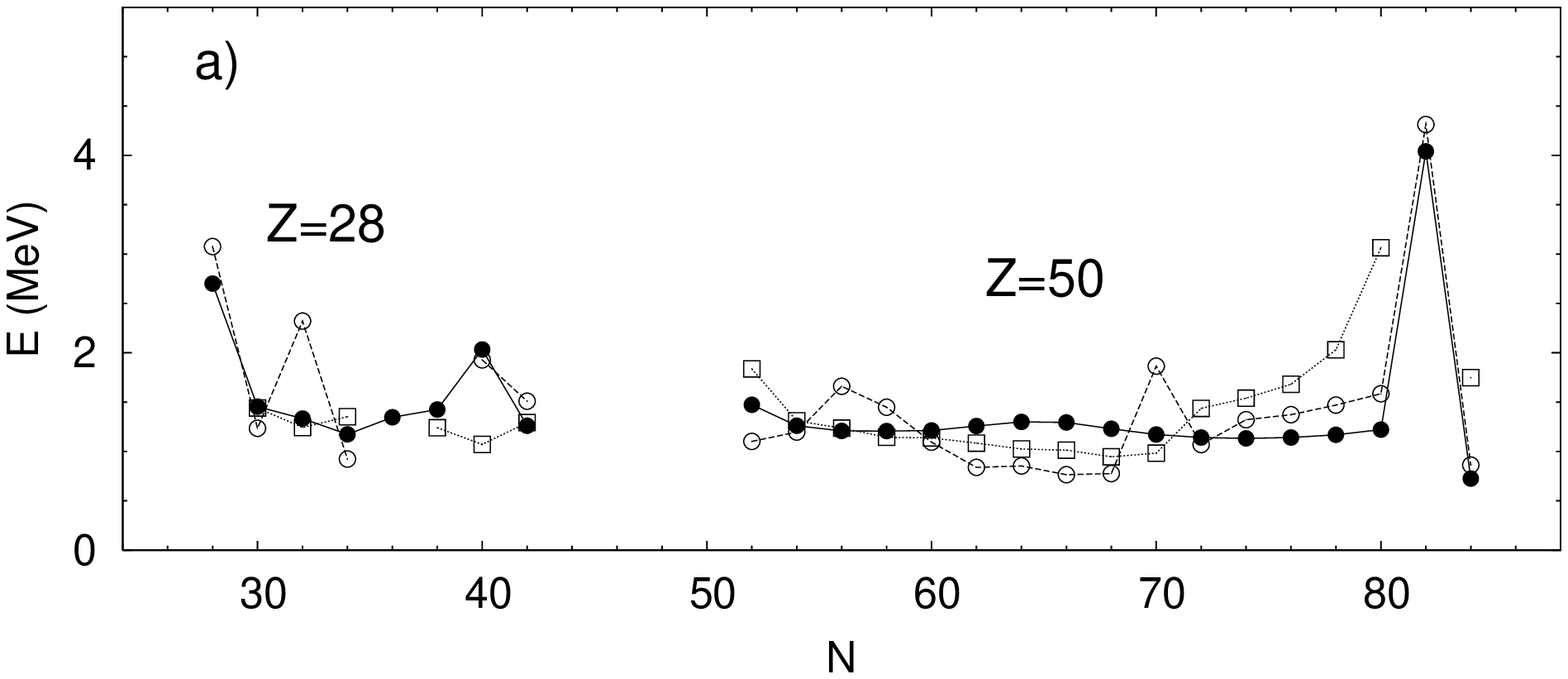}\\[.3cm]
\includegraphics[height=4.3cm]{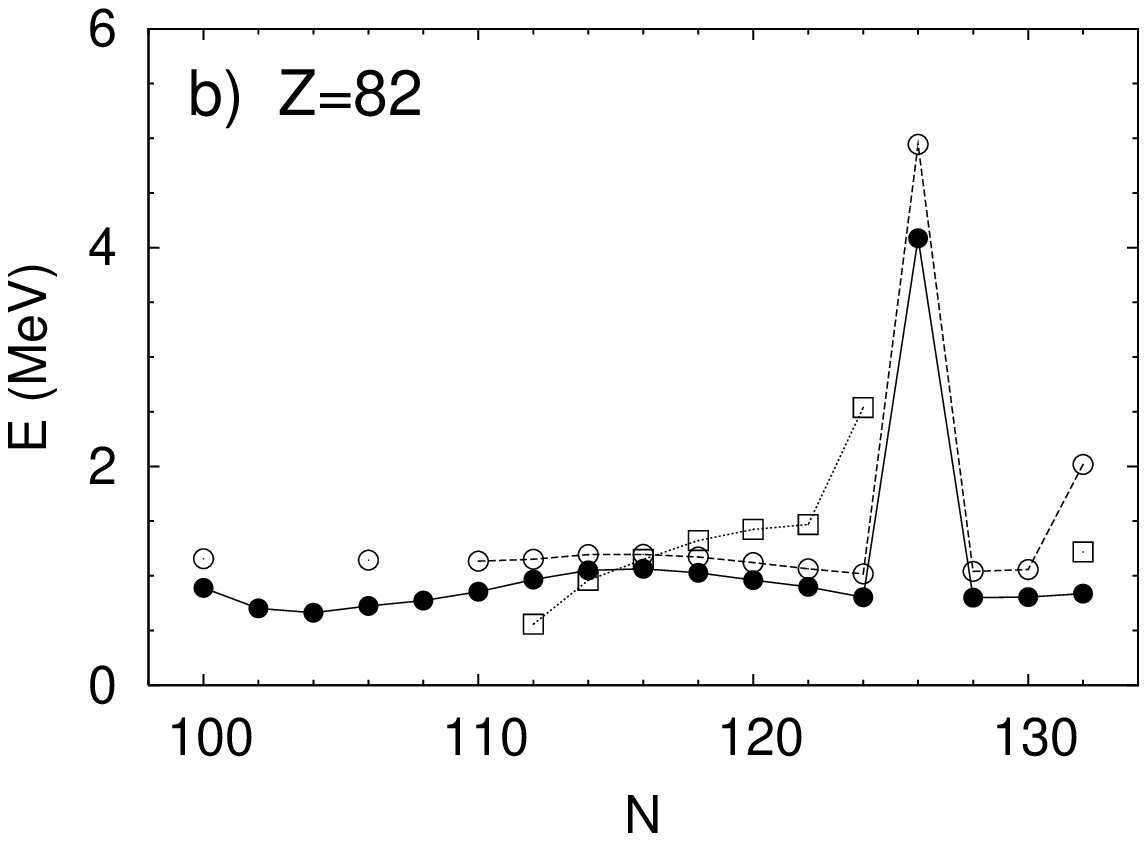}
\includegraphics[height=4.3cm]{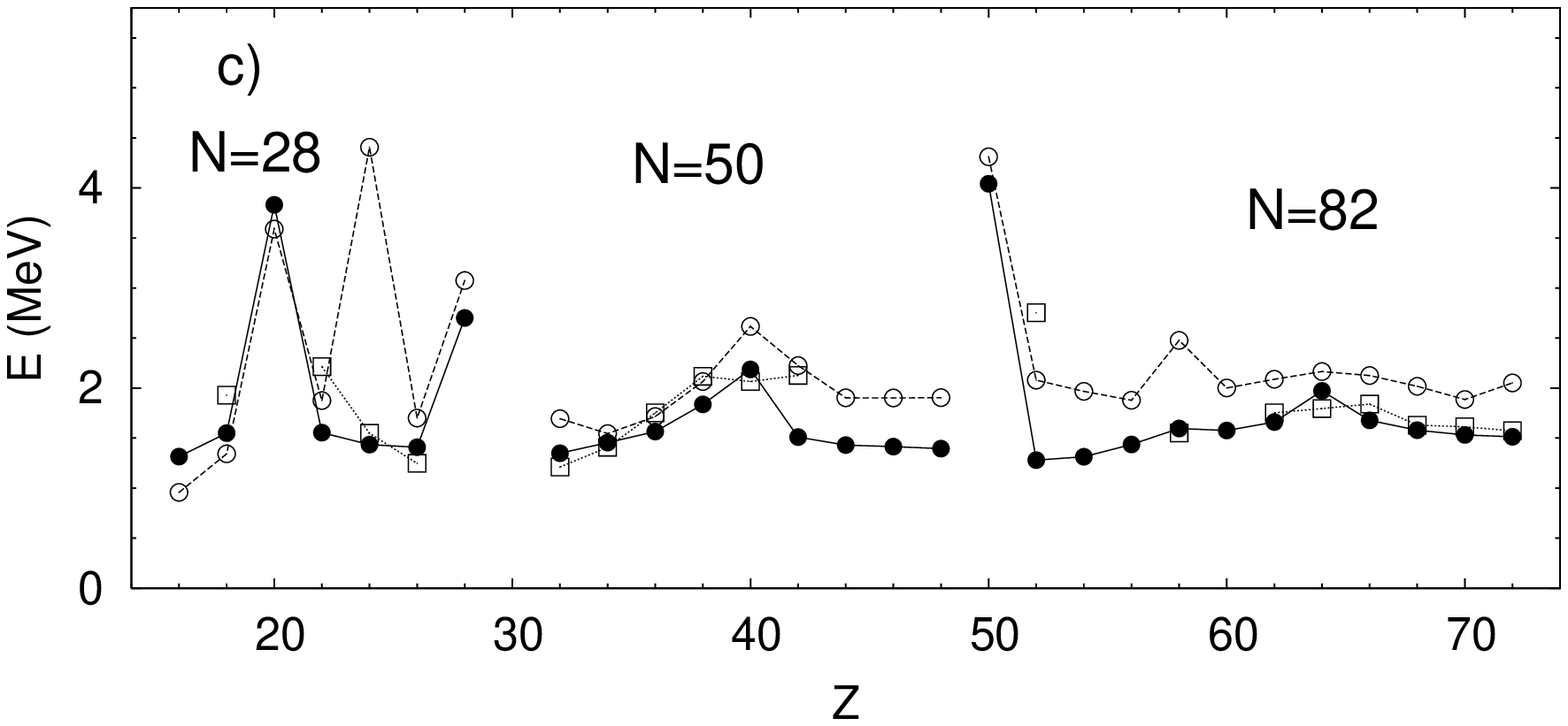}
\caption{\label{fig:isotopes}  The lowest $2^+$ energies of even 
Ni, Sn, and Pb isotopes (panels a and b), 
and isotones with $N=$ 28, 50, and 82 (panel c).  The open circles represent 
our results (SkM$^\ast$), the
squares the GCM-5DCH results of Ref.\ \cite{Gou07},
and the filled circles experimental data.}
\end{flushleft}
\end{figure}
in the second
line are the results from Ref.\ \cite{Gou07} for the Gogny functional.
We see that the QRPA (SkM$^*$) has a somewhat better average energy.  On
the other hand, the fluctuation of the transition matrix elements
is much better described by the GCM-5DCH (Gogny) treatment. 

We turn finally to semimagic isotope chains.  In Figure \ref{fig:isotopes} we
show of our predictions (SkM$^\ast$) and those of Ref.\ \cite{Gou07}  (along with
experimental data) for energies along 3 isotopic chains --- $Z=28$, 50, and 82
--- and isotonic chains with the same magic numbers.  Here our calculation is
again systematically better than that of Ref.\ \cite{Sab07}, the results of
which we omit from the plot for the sake of clarity.  Our calculation and the
GCM-5DCH calculation of Ref.\ \cite{Gou07} have complementary virtues in the
isotopic chains.  The nearly unchanging energies for $Z=50$ ($52 \leq N \leq
80$) and $Z=82$ ($112 \leq N \leq 124$), used to motivate the idea of
generalized seniority \cite{Tal71}, have long challenged mean-field-based
approaches.  In $Z=50$, for example, the GCM-5DCH reproduces this trend quite
well up to $N=70$, but then incorrectly predicts a gradual increase in
excitation energy as the closed shell at $N=82$ is approached.  Our calculation,
by contrast, produces lower energies than experiment around $N=62$ because the
QRPA solutions are close to a transition to quadrupole deformation, but
accurately reproduces the sharp jump at $N=82$.  In the $Z=82$ chain we
reproduce the energies well for $112 \leq N \leq 124$, a bit better than does
the GCM-HW, particularly near the closed shell.  In the isotonic chains, by
contrast, the results of the GCM-HW are pretty uniformly better than ours.
Again, we cannot be sure how much of the difference is due to functionals and
how much to methodology.  Ref.\ \cite{an06}, which uses a relativistic functional
to examine some of these same chains in the QRPA, obtains results that appear
comparable to ours.

\section{Conclusion}
We have used the QRPA to calculate the energies and E2 strengths
of the lowest $2^+$ states in a wide range of even-even spherical nuclei, and
compared the results with experiment for more than 150 energies and
more than 80 strengths.  We applied two functionals, SLy4 and 
SkM$^*$. On the whole, SkM$^*$ performed better than SLy4. 
For energies,  our calculation with SkM$^*$ is comparable to the GCM-5DCH
calculation of Ref.\ \cite{Gou07} in spherical nuclei ---
better near closed shells, though perhaps not quite as good at midshell ---
and better than the GCM-HW calculation of Ref.\ \cite{Sab07}.
For $B(E2)\!\!\uparrow$ values, the GCM-5DCH calculations appear to
be the best if doubly-magic nuclei are excluded.   The QRPA is a small
amplitude approximation, and its results are not disappointing, considering
that limitation. 

We were surprised to find that the two functionals disagreed significantly on
the question of which nuclei are spherical, an issue that arises at the
mean-field level and has nothing to do with the QRPA.  We plan to broaden our
study soon to include deformed nuclei, allowing a more comprehensive comparison
of approaches and functionals.

\section{Acknowledgment}
We thank A. Bulgac for discussions.  This work was supported by the UNEDF SciDAC
Collaboration under DOE grant DE-FC02-07ER41457.
We used computers at the National Energy Research Scientific Computing Center
at Lawrence Berkeley National Laboratory.

\appendix

\section{The RPA and the GCM} In this appendix, we present a simple
model that indicates why the GCM overpredicts the energy of the lowest 2$^+$ excitation
under conditions for which the RPA works well.    The effect of the constraining
field $Q$ on the mean-field ground state $|0\rb$ can be expressed in
perturbations theory (that is, for small deformation) as 
\begin{equation} |q\rb = |0\rb - \sum_i {| i\rb \lb i|Q|0\rb\over
E_i} \label{eq:q} 
\end{equation}
where $|0\rb$ is the spherical static-DFT state, $|i\rb$ the RPA
eigenstate, and $E_i$ the corresponding RPA energy.  In the GCH-HW approach, 
one first determines the polarized state $|q\rb$.  The state is then projected
so as to have a well-defined 
angular momentum.  The excitation energy is taken to be the
expectation value of the Hamiltonian in the projected state.  In
Eq.~(\ref{eq:q}), the projection eliminates $|0\rb$ but keeps the relative
amplitudes of the excited components in the above wave function.  The
expectation value of the Hamiltonian is then given by 
\begin{equation} \lb E\rb_Q = {  \sum_i
{\lb i|Q|0\rb^2 \over E_i} \over \sum_i {\lb i|Q|0\rb^2\over E_i^2} } 
\end{equation}

To get 
a quantitative estimate of error in the GCM-HW energy in this small-amplitude
limit, we make a two-state approximation to the
RPA spectrum, which in fact in spherical nuclei is often dominated by two
strong transitions: an in-shell transition at a few
MeV or less and the giant isovector quadrupole resonance at 10$-$15 MeV.  The properties we
need are their transition strengths and the ratio of their energies
$r=E_1/E_2$.  It is convenient to express the transition strengths as a
fraction of the energy-weighted sum rule: 
\begin{equation} 
f_i = {E_i \lb0| Q| i\rb^2\over
\sum_j E_j \lb 0| Q| j\rb^2}\,.
\end{equation}  
With some simple algebra the energy
expectation value $\lb E \rb_Q$ can be written in terms of the lowest 
RPA energy $E_1$
as \begin{equation} \lb E \rb_Q = E_1 \left({ f_1  + (1-f_1) r^2\over f_1 +(1-f_1)r^3}
\right) \label{eq:err} 
\end{equation}
%Given that $f_1$ is a small fraction, in the range $0.05-0.25$, 
The right hand
side approaches the correct value ($=E_1$) as $r \rightarrow 0$.  

One case for which the small amplitude approximation (and therefore the RPA)
should work well is the doubly magic nucleus \pb.  In our QRPA calculation with
the SLy4 functional, we find an excitation energy of $5.0$ MeV with a transition
strength that amounts to 17\% of the isoscalar sum rule.  The giant resonance is
located at $E_2\approx 12$ MeV, giving $r \approx 0.4$.  Inserting these numbers
in Eq.\ (\ref{eq:err}) results in a GCM-HW energy that is 37\% too large, and is
close to the value obtained in Ref.\ \cite{Sab07}, $E_2(\textrm{GCM}) = 6.8$
MeV.  This nucleus and others near closed shells, it should be noted, are not
good ones for the GCM because of the size of $r$.  Away from doubly magic nuclei
$r$ is smaller, resulting in a smaller GCM error for a given $f_1$.  

It might be possible to improve the GCM by adding an additional quadrupole field
that would distinguish the giant resonance from the low-lying collective states.  

\section{Sum-rule fraction}
It has been known for a long time that the \twop excitation energies
and the associated $B(E2)$$\uparrow$ values have a strong inverse 
correlation\cite{gr62}, and it is interesting to see how well the
theoretical results reproduce this behavior.  The product $E_2 B(E2)$$\uparrow$
can be expressed as a fraction of an energy-weighted sum rule, 
giving it the $A$-dependence of the sum rule.  A form that is often used is $ Z^2 A^{-\alpha}$, where
$\alpha=1/3$ corresponds to the isoscalar sum rule for a liquid
drop\cite{BM2}.  We show in Table \ref{tab:sumrule} the average of the scaled
product
\be
\label{eq:S}
S= {E_2 B(E2)\!\!\uparrow \over Z^2 A^{-1/3}}
\ee
obtained from averaging $R_S= \log(S)$.  The variance of $R_S$ is also
given in the table.  The first 
line shows the values obtained from the experimental data  \cite{Ram01}.  The second line
contains only the nuclei predicted to be spherical with the SkM$^*$
functional.  One sees that the average over the spherical nuclei is close
to the global average, indicating that the sum-rule fraction is
insensitive to deformation.  The variance is larger among
the spherical nuclei, however.  The third line shows the results obtained in the QRPA
with the SkM$^*$ functional.  The theoretical average value is 
significantly smaller than the empirical value.  Since the fraction of
the sum rule is a measure of the collectivity, it may be that our
functional does not have sufficient collectivity.  The sum-rule fraction
might be sensitive to the strength of the pairing interaction or some
other aspect of the functional, and it would be interesting to 
investigate that issue further.  

The
table also shows that the fluctuation in the sum-rule fraction, $\sigma_S$,
is significantly larger in the DFT calculation than in experiment.
Shell structure may need to be suppressed 
in some way, perhaps by increasing the pairing strength.

\begin{table}
\caption{\label{tab:sumrule}
Average sum rule fraction (Eq.\ (\ref{eq:S}) ) and  variance 
}
\begin{ruledtabular}
\begin{tabular}{lrccc}
source     & data set & Number of nuclei&  $\overline S= \exp(\overline{R}_S)$ 
&  $\sigma_S$\\
\hline
experiment   & all &  328 &$5.4 ~10^{-4}$ & 0.46\\
experiment  & spherical & 99 &$4.7 ~10^{-4}$  & 0.63\\  
DFT  & spherical  & 99& $2.8 ~10^{-4} $ & 0.95 \\    
\end{tabular}
\end{ruledtabular}
\end{table}

\end{document}